\documentclass{aa}  

\usepackage{graphicx}
\usepackage{txfonts}
\usepackage[colorlinks=true,allcolors=blue]{hyperref}
\usepackage{orcidlink}
\usepackage{booktabs}
\usepackage{float}

\begin{document}

   \title{
   What does it mean to take the mean?
   }

\authorrunning{Konstantinou et al.}
\titlerunning{}

   \subtitle{The effect of the averaging scale on the characterization of interstellar turbulence}

   \author{A. Konstantinou
           \inst{1,2,3 \orcidlink{0000-0002-4758-212X}}\fnmsep\thanks{akonstantinou@physics.uoc.gr}, 
           E. Ntormousi
           \and
           \inst{2 \orcidlink{0000-0002-4324-0034}}\fnmsep\thanks{evangelia.ntormousi@sns.it}, 
           K. Tassis\inst{1,3 \orcidlink{0000-0002-8831-2038}}
          }

   \institute{Department of Physics, University of Crete, Voutes Campus, GR-70013, Heraklion, Greece
         \and Scuola Normale Superiore, Piazza dei Cavalieri, 7 56126 Pisa, Italy
         \and  Institute of Astrophysics, Foundation for Research and Technology-Hellas, Vasilika Vouton, GR-70013 Heraklion, Greece
             }

   \date{Received ; accepted }

 
\abstract
  {
  In interstellar medium studies, defining an ordered and a random velocity or magnetic field extends beyond the scope of simple thought experiments. It often plays a crucial role in interpreting the dynamics of turbulence, both in simulations and in observations.
  }
  {We investigate how the choice of averaging scale affects the measurement and characterization of magnetic and kinetic turbulence in Milky Way-sized disk galaxies with different initial magnetic morphologies.
  }
  {We analyze two magnetohydrodynamic simulations of isolated disk galaxies, one initialized with a toroidal magnetic field (Model T) and the other with a random field (Model R). Using a spherical filtering method, we decompose the magnetic and velocity fields into mean and fluctuating components while varying the averaging scale
  and examine their energy ratios and power spectra as functions of time and averaging radius.
  
  }
  {Both models develop ordered and turbulent magnetic structures, whose relative strengths vary strongly with the averaging radius. The power spectra of the velocity and magnetic mean fields steepen with increasing smoothing scale, tracing the transition from coherent to turbulent regimes. The turbulent kinetic energy dominates over the magnetic counterpart, though the latter remains dynamically significant. Very importantly, we find that these ratios depend more strongly on the averaging radius than on the initial conditions of the magnetic field.
  }
  {The characterization of turbulence as strong or weak, 
  meaning whether or not fluctuations dominate over the mean depends sensitively on the chosen averaging scale, rather than being an intrinsic property of the system. The strong dependence of the turbulent fractions on the averaging scale has direct implications for magnetic field estimates obtained from observational methods such as the Davis-Chandrasekhar-Fermi technique. Careful consideration of scale is therefore essential when interpreting magnetic and kinetic turbulence in galaxies.
  
  }

\maketitle
%

\section{Introduction}
\label{sec:introduction}
 
Magnetic fields play a crucial role in galaxy evolution, because they can influence star formation \citep[see, e.g.,][for recent reviews]{hennebelle,pattle}, shape the dynamics of interstellar gas (e.g., by modulating disk pressure, affecting cloud fragmentation, and setting the disk vertical structure \citep{girichidis18,shykurov18}), 
and regulate the propagation of cosmic rays through the disk and halo \citep{fermi,cesarsky,desiati,shukurov}. Since magnetic fields are frozen into the interstellar plasma, their morphology reflects the interplay between coherent large-scale flows and small-scale turbulence, providing clues about the origin of fluctuations. The distinction between ordered and turbulent components is therefore central to understanding their role in galaxy evolution.

Dynamo theory is one field where the definition of a mean field is central.
Large-scale dynamos, driven by differential rotation and helical turbulence, generate coherent patterns extending over kiloparsec scales, whereas small-scale dynamos amplify random fields on the scale of the turbulent driving \citep[see][for a recent review]{brand_ntorm23}. In practice, the relative balance between these ordering and disordering mechanisms is shaped by galactic processes: differential rotation and shear can sustain large-scale dynamos, while turbulence from supernova feedback and other local instabilities can disrupt coherent structures and drive random fields. The resulting morphology of galactic magnetic fields thus reflects the competition between these mechanisms as well as the imprint of the initial field configuration, whose role remains comparatively unexplored.

Observationally, disentangling ordered and turbulent magnetic components remains a major challenge. For example, the total synchrotron emission measures the overall magnetic field projected on the plane of the sky, but it does not distinguish between coherent and random components. Polarized synchrotron emission, on the other hand, traces only the ordered part of the field, since contributions from isotropic turbulence cancel out in polarization. \citep{ferriere,beck,lopez}. However, instrumental effects such as beam smearing, interferometric filtering, and Faraday depolarization complicate this separation by mixing scales and suppressing polarized signals. \cite{hu_lazarian}, for example, show how the removal of low spatial frequencies in interferometric data alters the inferred magnetic field morphology. Large-scale surveys
such as Planck polarization maps of the Milky Way \citep{planck16} and the SALSA program for nearby galaxies \citep[][]{lopez,Borlaff_2023}, provide powerful constraints but also highlight these observational limitations. In this context, numerical simulations offer a complementary approach, allowing us to decompose magnetic fields into mean and fluctuating components in a controlled manner and directly explore  
the implications of these definitions.

In theoretical and numerical studies, several strategies have been developed to define the mean field. Since ensemble averaging is not possible even in numerical simulations due to the very high cost of repeating the experiment, a common approach is to separate the magnetic field into mean and fluctuating components by averaging over the entire computational domain, or, in systems stratified along the vertical ($z$) direction, by averaging in horizontal $(x,y)$ planes \citep[“horizontal averaging”; see][]{brandenburg2005}. These methods yield mean fields that are perfectly uniform or depend only on $z$. Although powerful in some contexts, such definitions may not fully reflect the spatial complexity expected in an inhomogeneous medium such as the ISM.

To capture more of this complexity, Gaussian smoothing has been introduced as an alternative.
\citet{hollins} applied this technique in three-dimensional MHD simulations of a $1 \times 1 \times 2$ kpc shearing box to extract mean magnetic, density and velocity fields, as well as magnetic and kinetic energy densities. From spectral analysis, they found that a smoothing length of approximately 75 pc effectively separates large-scale ordered structures from small-scale fluctuations in their specific setup, while emphasizing that this scale is not universal and may depend sensitively on the simulation parameters. Their results also showed that Gaussian smoothing, unlike horizontal averaging, is capable of preserving large-scale three-dimensional features of the mean field. A similar approach was adopted by \citet{gent13}, who employed Gaussian kernel averaging in stratified differentially rotating disk simulations driven by supernova turbulence. Their analysis revealed that the mean and fluctuating fields exhibit different growth rates and distinct integral scales (about 0.7 and 0.3 kpc, respectively), providing clear evidence of scale separation between ordered and turbulent components. These studies have demonstrated the value of scale-dependent filtering techniques in mean-field analysis.

In addition to horizontal averaging and Gaussian smoothing, other filtering approaches have been proposed to study how the averaging scale affects the inferred properties of turbulence.
For example, multi-point structure functions, such as those applied by \cite{lee2025} in observations of the Small Magellanic Cloud, can identify the characteristic driving scales of turbulence.
As discussed by \cite{cho2019}, a three-point structure function is mathematically equivalent to a ring-like filter in 2D or a spherical shell-like filter in 3D. These techniques provide complementary ways to separate large- and small-scale fluctuations, although their effectiveness depends on the presence of a dominant driving scale, which may not be clearly defined in systems with broad, multiscale turbulence like galactic disks.
Moreover, in principle, they are not immune to instrument-related averaging.

In this work, we take a complementary approach by applying a spherical averaging method with a variable radius in disk galaxies. The mean field is defined as the average within spheres of different sizes, which lets us examine how the balance between ordered and turbulent components depends on the averaging scale.
Using this framework, we analyze the time evolution and the power spectra of the magnetic and velocity fields in two Milky-Way-sized disk galaxy models that differ only in their initial magnetic morphology: one with a large-scale toroidal field and one with a random field, both with the same strength and exponential radial/vertical decline.
This experiment is particularly interesting because, due to magnetic flux conservation, we might expect the signature of the ordered initial conditions to persist as a mean field. Therefore, this setup allows us to isolate the role of initial conditions while exploring three key questions: How does turbulence strength depend on the averaging scale and initial field configuration? How does the inferred degree of magnetic ordering vary with the spatial scale used to define the mean field? What processes drive the transitions between ordered and random magnetic fields?

We describe the numerical code and the setup in Sect.~\ref{sec:methods}. We present the results of our investigations in Sect.~\ref{sec:results}, a discussion in Sect.~\ref{sec:discuss}, and the conclusions in Sect.~\ref{sec:conclus}.
\section{Method and setup}\label{sec:methods}

\subsection{Numerical simulations}

We analyze two galaxy-scale MHD simulations originally published in \cite{konstantinou} (hereafter K24). Here we outline the most relevant features of the models and refer the reader to K24 for details.

The initial conditions of the two galaxies are identical, except for the magnetic-field morphology. Model T is initialized with a large-scale toroidal magnetic field, while model R begins with a random magnetic field. In both cases, the magnetic field strength follows an exponentially declining profile.

\subsection{Turbulence driving}

Turbulence in our simulations is driven by both supernova feedback and the differential rotation of the galactic disk. Supernova explosions inject energy locally into the ISM through expanding shells and shocks, while differential rotation continuously drives large-scale solenoidal motions through shear.

Although supernovae are essential for shaping the multiphase structure of the ISM, the global turbulent energy budget in our models is dominated by large-scale shear associated with differential rotation. As a result, turbulent energy injection occurs primarily on large spatial scales, with additional contributions from supernova feedback over a broad range of smaller and intermediate scales. 
However, other fluid instabilities are almost certainly acting on different scales and all the unstable modes can also interact non-linearly, so it is not clear that the main turbulence drivers we mention above can be clearly identified in post processing.

\subsection{Definition of mean and turbulent field}
To separate the mean and turbulent components of the magnetic field, we adopted a spherical averaging (filtering) procedure. Since the data are obtained from an adaptive mesh refinement (AMR) simulation, the spatial resolution varies, and therefore it is not feasible to define the neighborhood of each cell using a fixed number of cells. Instead, for each grid cell, we define a sphere centered on the cell and compute the average magnetic field within spheres of different radii: 0.1, 0.5, and 1 kpc. The turbulent component is then obtained by subtracting this local mean field from the total field. This method provides a flexible way to examine how the definition of the mean field depends on the spatial scale, capturing local variations while preserving large-scale structures.

This framework enables a systematic study of the time evolution and power spectra of both the mean and turbulent magnetic fields. By comparing models T and R, we can explore how different initial magnetic configurations influence the development of ordered and random components across multiple spatial scales.

To validate our procedure, we first applied it to a set of test cases. For a uniform magnetic field, our method correctly recovered the mean field as identical to the total field, with the turbulent component vanishing ($\delta B = 0$). We then tested a purely toroidal configuration and again obtained the expected result: The mean field matched the total field without fluctuations. Finally, we introduced random perturbations corresponding to 20\% of the background field. The method accurately reproduced the imposed level of turbulence, yielding $\delta B / B_{0} \approx 0.2$, respectively, as shown by the histograms in Appendix~\ref{appendix:a}.

\section{Results}\label{sec:results}

\subsection{Maps of the magnetic and velocity fields}

\begin{figure*}[!ht]
    \centering
    \hspace{-1.2cm} \large Model T  \hspace{5.7cm} \large Model R \\
    \rotatebox{90}{\hspace{3.5cm} \Large$B_{tot}$}
    \includegraphics[trim={0cm 0cm 3.8cm 1cm},clip,width=.44\linewidth]{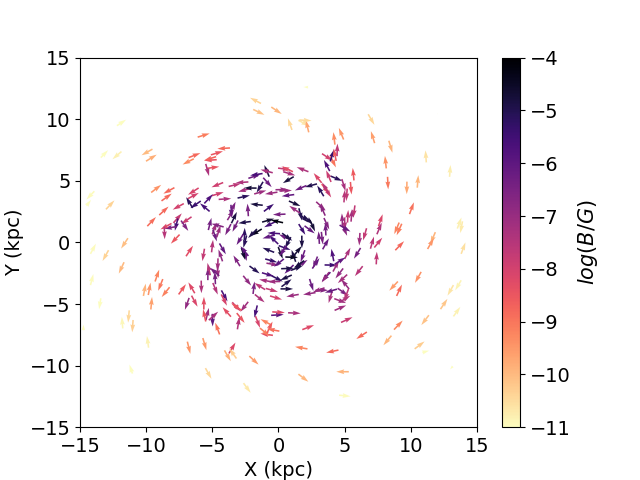} 
    \includegraphics[trim={1.8cm 0cm 0cm 1cm},clip,width=.51\linewidth]{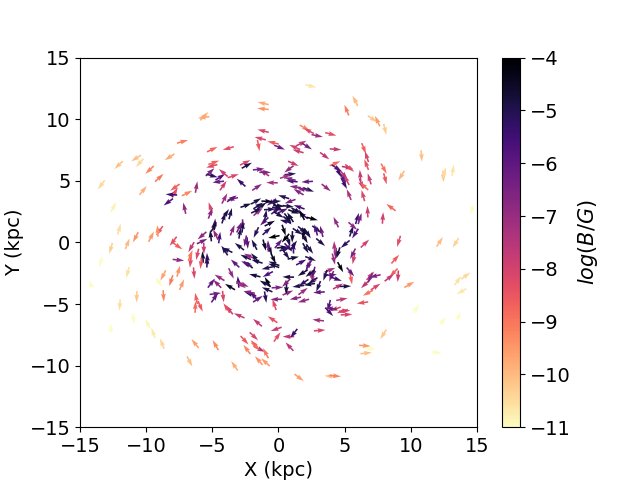}\\
    \rotatebox{90}{\hspace{3.5cm} \Large$B_{0}$}
    \includegraphics[trim={0cm 0.cm 3.8cm 1cm},clip,width=.44\linewidth]{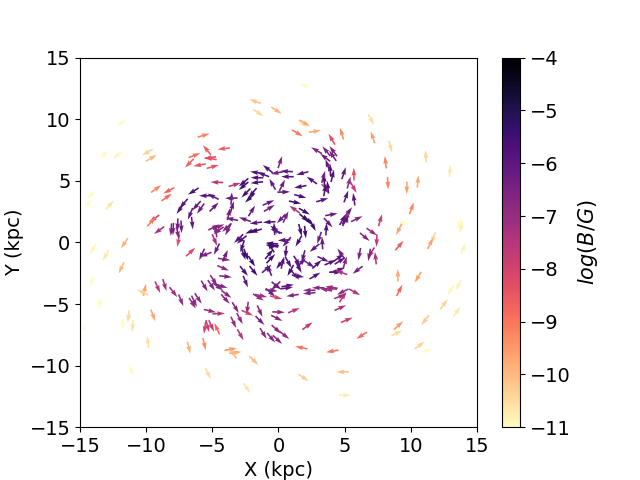} 
    \includegraphics[trim={1.8cm 0cm 0cm 1cm},clip,width=.51\linewidth]{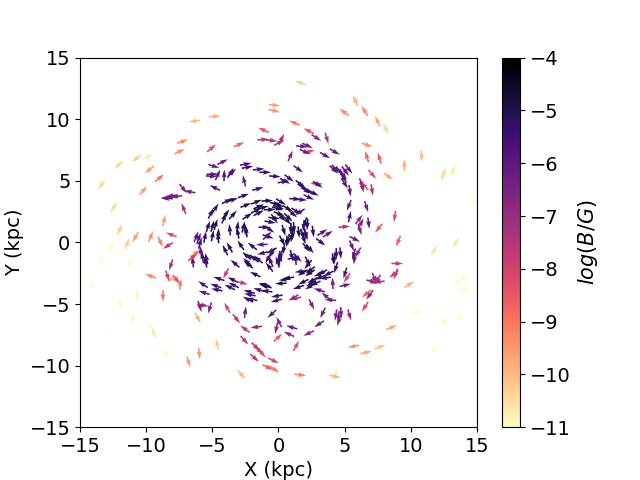}\\
    \rotatebox{90}{\hspace{3.5cm} \Large$\delta B$}
    \includegraphics[trim={0cm 0cm 3.8cm 1cm},clip,width=.44\linewidth]{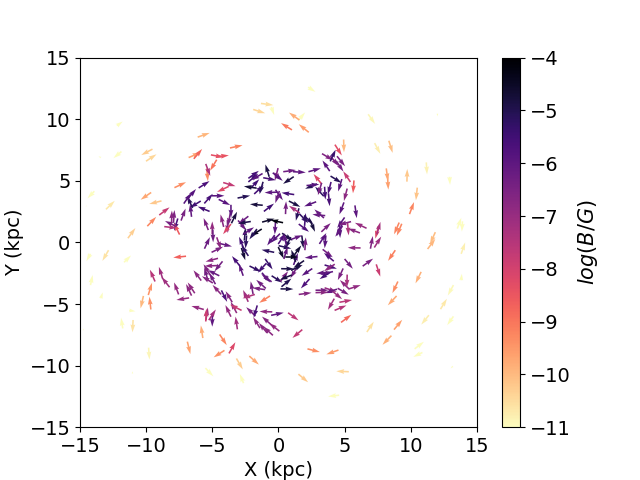} 
    \includegraphics[trim={1.8cm 0cm 0cm 1cm},clip,width=.51\linewidth]{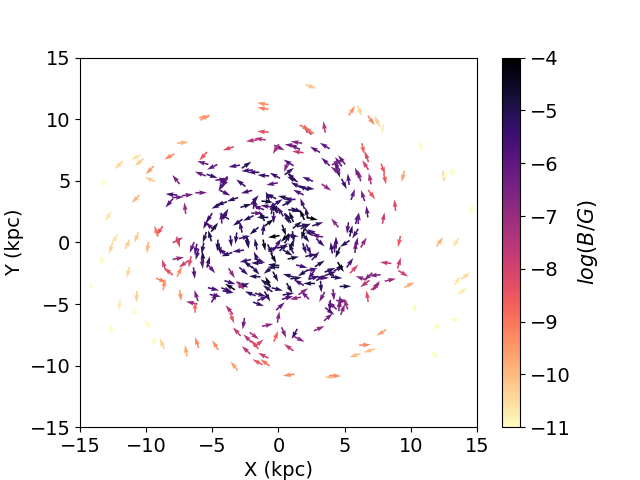}
    \caption{Face-on maps of the projected magnetic field vectors for the total field ($B_{tot}$), the mean field ($B_0$) and the turbulent component ($\delta B$) at 500 Myr. Model T and model R are shown on the left and right side, respectively. }
    \label{fig:B_maps_500}
\end{figure*}

Fig.~\ref{fig:B_maps_500} shows face-on maps of the projected magnetic field vectors for the total field ($B_{tot}$), the mean field ($B_0$) and the turbulent component ($\delta B$) at 500 Myr with an averaging radius of 1 kpc for model T and model R.
In the projected maps of the mean and turbulent magnetic fields, we find that the mean component exhibits clear large-scale ordered structures, as expected. Interestingly, the turbulent component also displays patches of apparently ordered patterns, in some cases with orientations opposite to the mean field. This arises naturally from the definition of the decomposition: the turbulent field is computed relative to a local average, and if the averaging scale does not fully capture the global field organization, part of the coherent structure can leak into the fluctuating component. Moreover, turbulence in astrophysical plasmas is not entirely random but often contains correlated or filamentary structures that can produce ordered signatures within the fluctuating field. Projection effects may further amplify this appearance by aligning fluctuations along the line of sight.

Although Model T begins with a globally ordered magnetic configuration, the maps show that Model R appears more ordered at later times. This likely results from the different spatial distribution of star formation and feedback activity \citep[see][]{konstantinou}.

\begin{figure*}[!ht]

    \centering
    \hspace{-1.2cm} \large Model T  \hspace{5.7cm} \large Model R \\
    \rotatebox{90}{\hspace{3.5cm} \Large$v_{tot}$}
    \includegraphics[trim={0cm 0cm 3.8cm 1cm},clip,width=.44\linewidth]{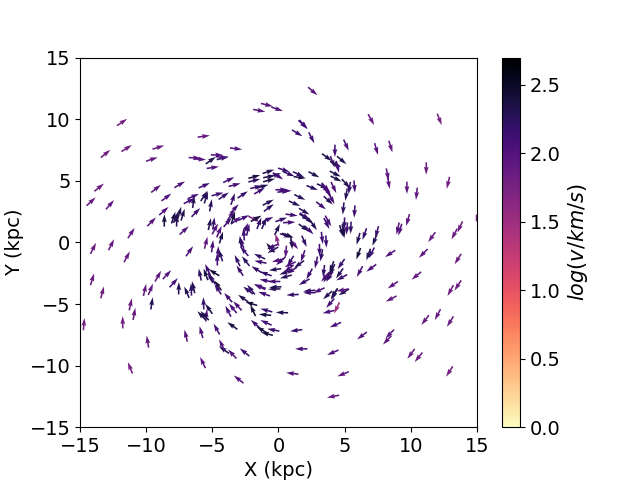} 
    \includegraphics[trim={1.8cm 0cm 0cm 1cm},clip,width=.51\linewidth]{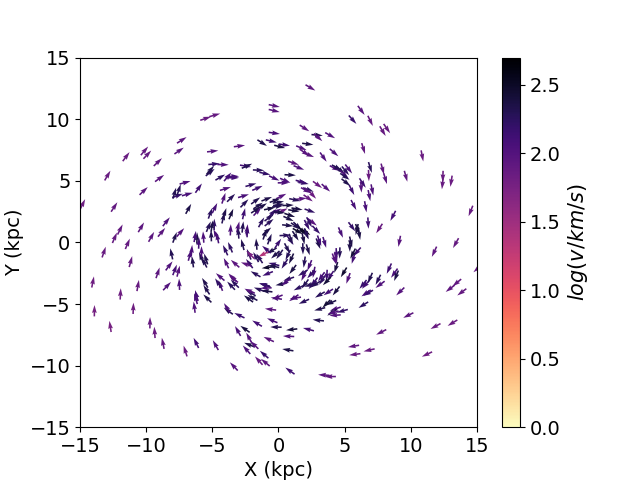}\\
    \rotatebox{90}{\hspace{3.5cm} \Large$v_{0}$}
    \includegraphics[trim={0cm 0.cm 3.8cm 1cm},clip,width=.44\linewidth]{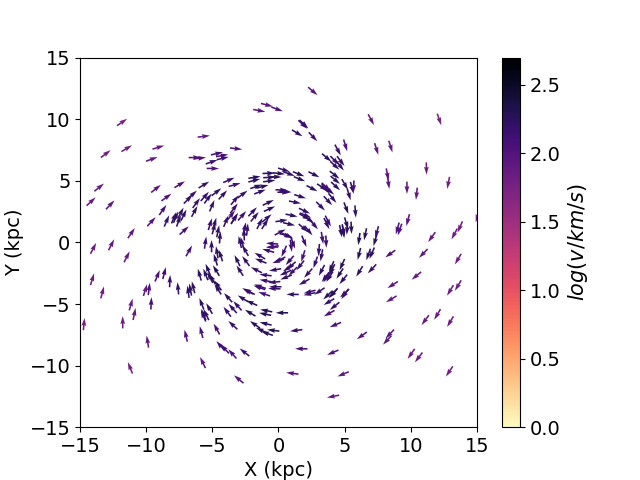} 
    \includegraphics[trim={1.8cm 0cm 0cm 1cm},clip,width=.51\linewidth]{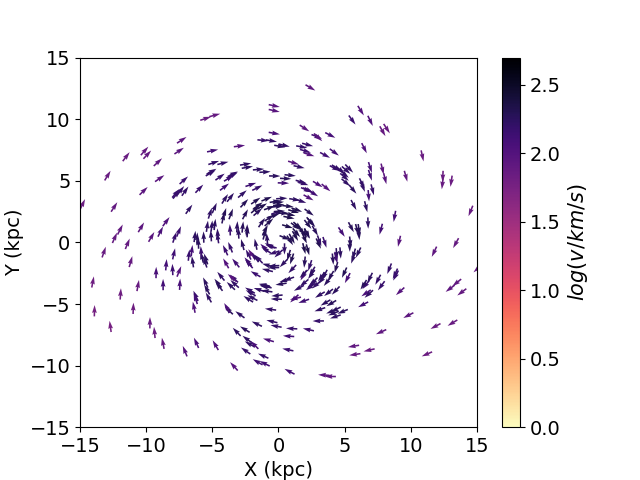}\\
    \rotatebox{90}{\hspace{3.5cm} \Large$\delta v$}
    \includegraphics[trim={0cm 0cm 3.8cm 1cm},clip,width=.44\linewidth]{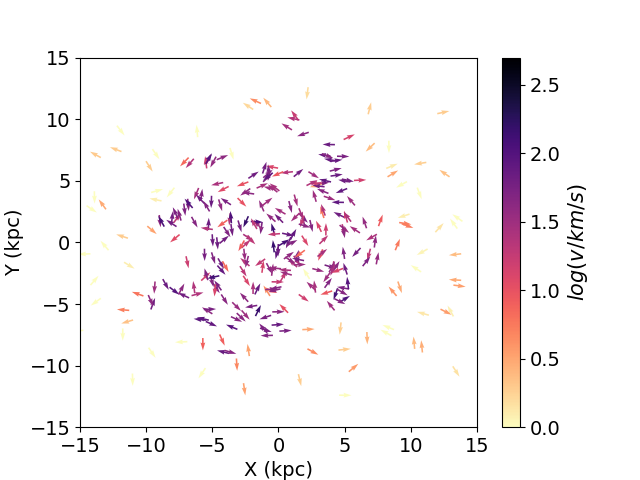} 
    \includegraphics[trim={1.8cm 0cm 0cm 1cm},clip,width=.51\linewidth]{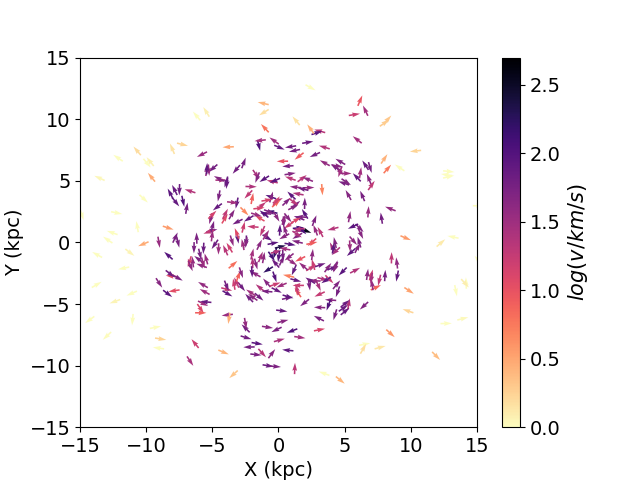}
    \caption{Face-on maps of the velocity field vectors for the total field ($v_{tot}$), the mean field ($v_0$) and the turbulent component ($\delta v$) at 500 Myr. Model T and model R are shown on the left and right side, respectively. }
    \label{fig:v_maps_500}
\end{figure*}

 Fig.~\ref{fig:v_maps_500} presents the corresponding maps for the velocity field. The mean velocity field ($v_0$) is mainly shaped by galactic rotation, which is expected given that these galaxies are dominated by dark matter and exhibit low star formation rates. Since the two models have identical initial conditions, except for the magnetic field morphology, their ordered velocity structures are expected to be similar. However, we notice some differences that indicate that the magnetic field can influence the gas dynamics, possibly through their interaction with density structures.
 The turbulent component ($\delta v$) highlights small-scale fluctuations induced by supernova feedback and local instabilities.

\subsection{Power spectra of the magnetic and velocity fields}

\begin{figure*}
    \centering
    \hspace{1.1cm}\large Model T  \hspace{6.9cm} \large Model R \\
    \includegraphics[trim={0cm 0cm 0.3cm 0cm},clip,width=.50\linewidth]{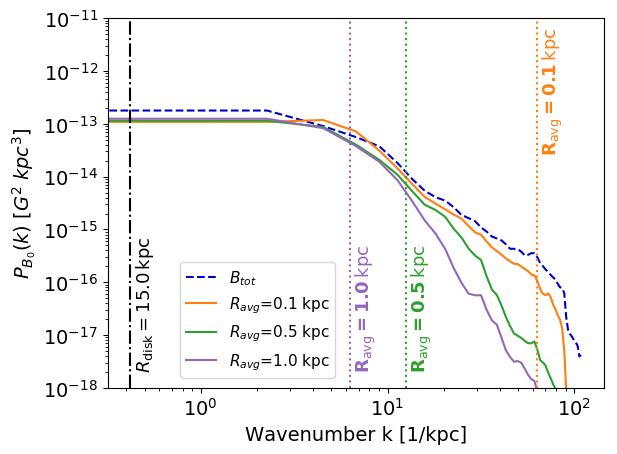} 
    \includegraphics[trim={0.86cm 0cm 0cm 0cm},clip,width=.48\linewidth]{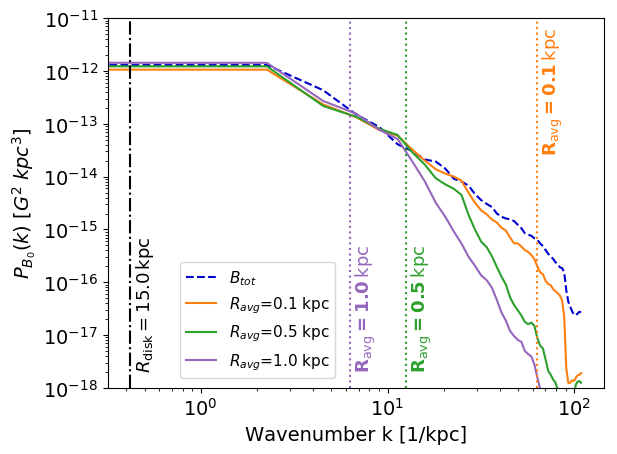}\\
    \includegraphics[trim={0cm 0.cm 0.3cm 0cm},clip,width=.50\linewidth]{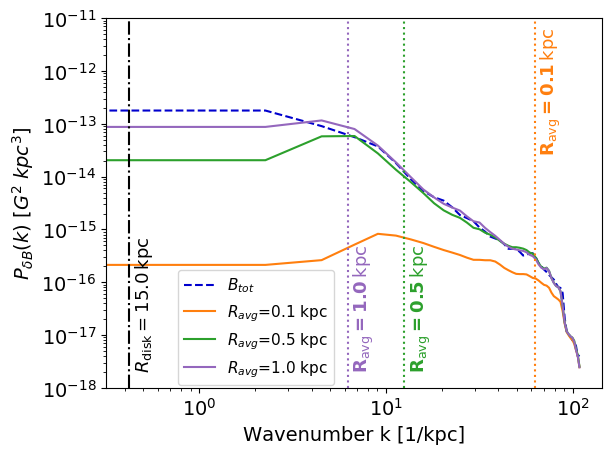} 
    \includegraphics[trim={0.86cm 0cm 0cm 0cm},clip,width=.48\linewidth]{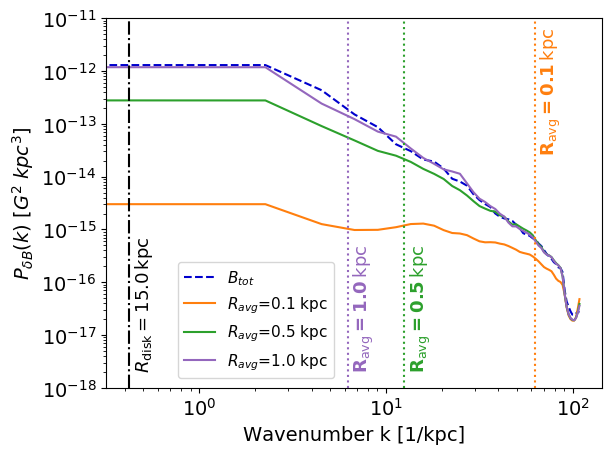}
    \caption{Power spectra of the mean ($P_{B_0}$) (top panel) and turbulent magnetic field ($P_{\delta B}$) (bottom panel) for different averaging radii at 500 Myr. The orange, green and purple correspond to $R_{avg}$=0.1 kpc, $R_{avg}$=0.5 kpc and $R_{avg}$=1.0 kpc. The dashed blue line represents the power spectrum of the total magnetic field. Model T and model R are shown on the left and right side, respectively.}
    \label{fig:ps_Β_500_2}
\end{figure*}

\begin{figure*}
    \centering
    \hspace{1.1cm}\large Model T  \hspace{6.9cm} \large Model R \\
    \includegraphics[trim={0cm 0cm 0.3cm 0cm},clip,width=.50\linewidth]{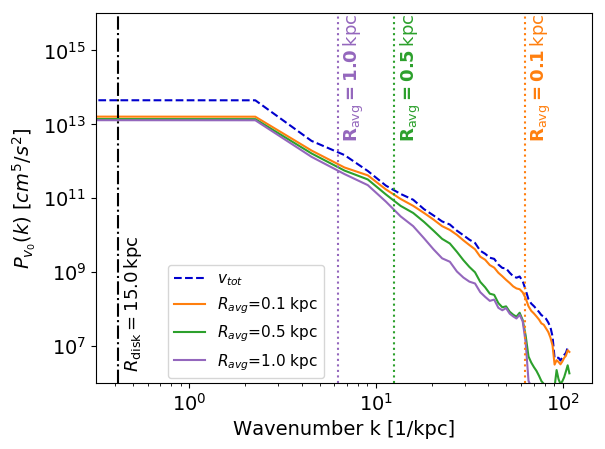} 
    \includegraphics[trim={0.86cm 0cm 0cm 0cm},clip,width=.48\linewidth]{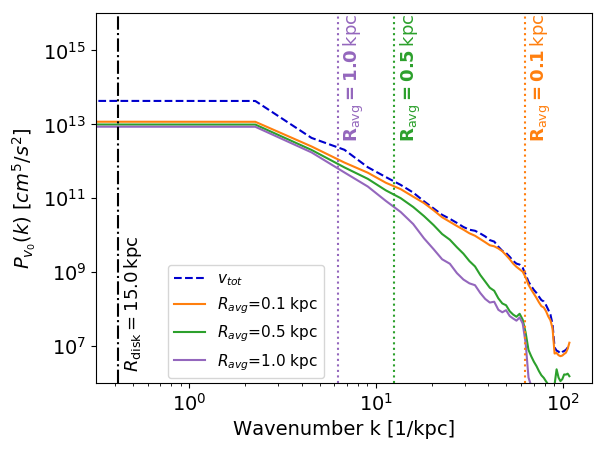}\\
    \includegraphics[trim={0cm 0.cm 0.3cm 0cm},clip,width=.50\linewidth]{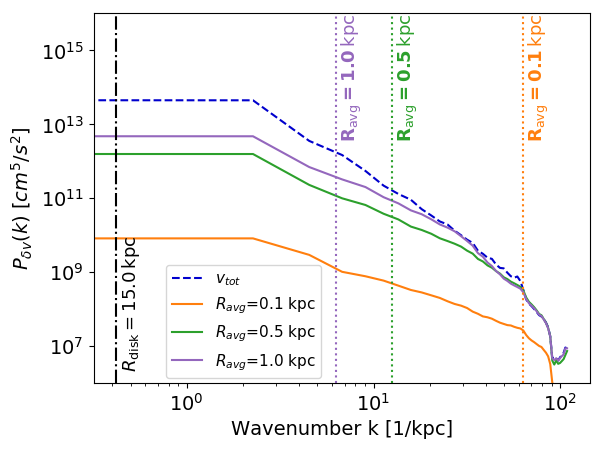} 
    \includegraphics[trim={0.86cm 0cm 0cm 0cm},clip,width=.48\linewidth]{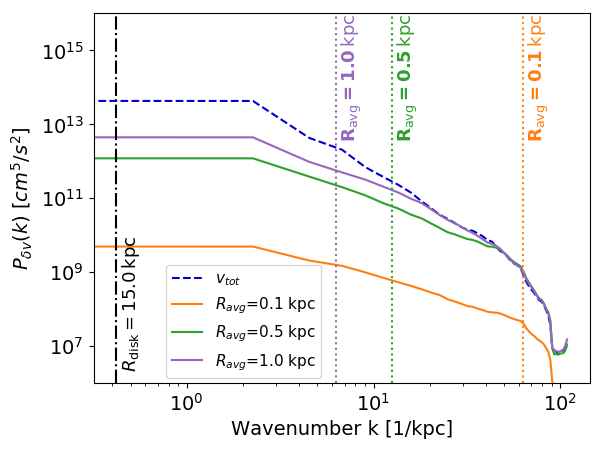}
    \caption{Power spectra of the mean ($P_{v_0}$) (top panel) and turbulent velocity field ($P_{\delta v}$) (bottom panel) for different averaging radii at 500 Myr. The dotted, dashed and solid lines correspond to $R_{avg}$=0.1 kpc, $R_{avg}$=0.5 kpc and $R_{avg}$=1.0 kpc. The dashed blue line represents the power spectrum of the total velocity field. Model T and model R are shown on the left and right side, respectively.}
    \label{fig:ps_v_500_2}
\end{figure*}

\begin{figure*}
    \centering
    \hspace{1.1cm}\large Model T  \hspace{6.9cm} \large Model R \\
    \includegraphics[trim={0cm 0.cm 0.3cm 0cm},clip,width=.50\linewidth]{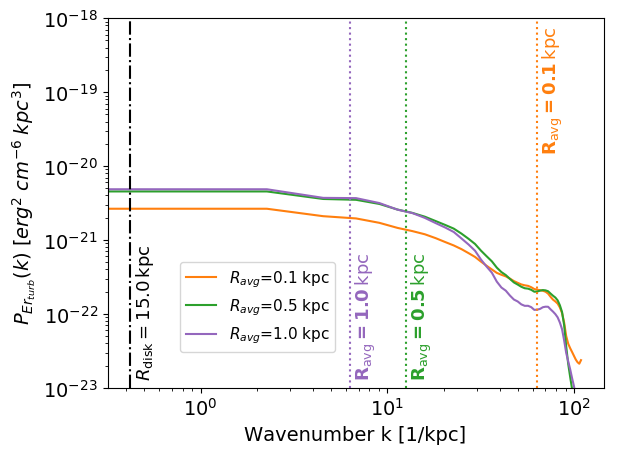} 
    \includegraphics[trim={0.86cm 0cm 0cm 0cm},clip,width=.48\linewidth]{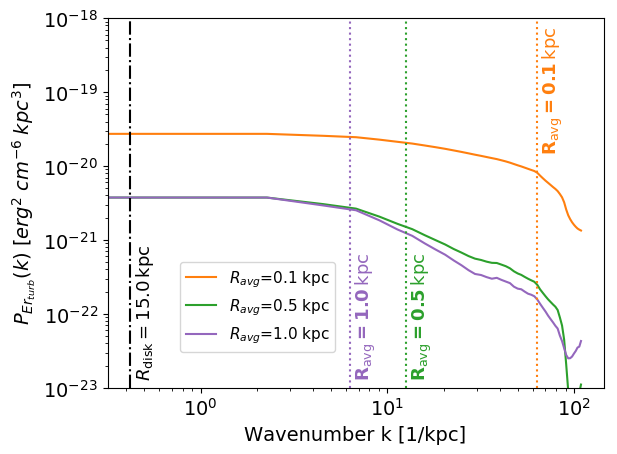}
    \caption{Power spectra of the residual energy density of the turbulent components ($P_{E_{r,turb}}$) for different averaging radii at 500 Myr. Model T and model R are shown on the left and right side, respectively.}
    \label{fig:ps_eres_500}
\end{figure*}

In Fig~ \ref{fig:ps_Β_500_2}, we analyze the power spectra of the total magnetic field ($B_{tot}$) and the mean field ($B_0$) for different averaging radii in the top panel, and of the fluctuating component ($\delta B$) in the bottom panel. For reference, we also indicate characteristic wavenumbers corresponding to the disk radius ($R_{disk}$) and the averaging radius ($R_{avg}$). 

For the mean magnetic field $B_0$, the power spectrum decreases sharply beyond the corresponding averaging radius, with the slope steepening from $-2.86$ at $R_{avg}=0.1$ kpc to $-5.57$ at $R_{avg}=1.0$ kpc in Model T, and similarly from $-3.39$ to $-6.16$ in Model R (Table~\ref{tab:slopes_by_radius} and Appendix~\ref{appendix:b}).  
This is expected since the mean field $B_0$ is calculated by averaging $B_{tot}$ on a finite spatial scale, 
fluctuations on scales smaller than the averaging radius are suppressed in $B_0$, leading to a loss of power at higher k in its spectrum. 

The turbulent component $\delta B$ exhibits a different trend. At small averaging radii ($R_{avg}=0.1$ kpc), the spectra are relatively flat, with slopes $-1.13$ (model T) and $-0.91$ (model R) for $k \in [12,60]$, but they steepen to $\sim-2.5$ at $R_{avg}=1.0$ kpc for both models. This reflects the increasing dominance of resolved turbulent fluctuations at larger averaging radii, once the mean field has been more heavily smoothed. Physically, this marks the transition from coherent, large-scale magnetic structures (captured in $B_0$) to small-scale turbulent fields ($\delta B$).
The large-scale power decreases for $R_{avg}=0.1$ kpc due to the local nature of the averaging.
The mean field follows the large-scale galactic structure, removing most of the coherent components from the fluctuations. For larger averaging radii, the mean field is smoother and less sensitive to local variations, so a larger fraction of the kpc-scale coherent magnetic structure is included in the turbulent component. This results in enhanced large-scale power in $\delta B$ and highlights the strong dependence of the magnetic field decomposition on the chosen averaging scale.

Both models exhibit these qualitative trends, but the overall power is higher in Model R across all scales. This behavior is expected since model T presents marginal amplification (A=1.2 \footnote{A is the ratio of the volume-weighted sum of the magnetic energy density at 500 Myr to that at 0 Myr.}) while model R undergoes slightly stronger amplification (A=4.57). Notably, the power spectra of the turbulent magnetic component in Model T display a broad peak in the spectrum at scales comparable to the disk size. This feature is not present in the velocity spectra and appears only in Model T, suggesting that it originates from the magnetic field configuration. In Model T, the initially ordered, disk-aligned magnetic field is amplified by shear and differential rotation, generating coherent perturbations at the disk scale that are captured as enhanced turbulent power. Model R, which starts from a more tangled magnetic configuration, lacks such organized shear-driven structures, leading to a smoother turbulent spectrum without a corresponding feature. The presence of this peak therefore represents an imprint of the initial magnetic field configuration in Model T.

Fig.~\ref{fig:ps_v_500_2} shows the power spectra of the velocity fields which show a similar trend: the power of the mean velocity field drops for scales smaller than the averaging radius, while the turbulent component dominates there. On large scales, the mean velocity reflects coherent bulk motions, such as galactic rotation, whereas on smaller scales, turbulence overtakes. Although the velocity power spectra of the two models are remarkably similar, minor differences are present. These variations are not unexpected at 500 Myr, given that both systems have undergone substantial nonlinear evolution with processes such as feedback. Overall, however, the velocity field remains largely governed by global disk dynamics and is less sensitive to the initial magnetic configuration than the magnetic field itself.

In Fig.~\ref{fig:ps_eres_500}, we show the power spectra of the residual energy, $E_r = E_{\text{kin}} - E_{\text{mag}}$, for the turbulent components. A clear dependence on the averaging scale is again evident, although the trend differs from that observed in the power spectra of the magnetic and velocity fields. For Model T, at $R_{avg}$ = 0.1 kpc, the residual energy spectrum exhibits the lowest amplitude among all radii, except at the smallest scales. In contrast, for Model R, the corresponding spectrum is significantly higher. Despite these differences, both models display the characteristic decrease in power at scales smaller than the averaging radius.

\begin{table}[!ht]
\centering
\caption{Power-law slopes of magnetic and velocity field spectra for Models T and R at different averaging radii ($R_{avg}$), fitted over the wavenumber range $k \in [12,60]$.}
\label{tab:slopes_by_radius}
\begin{tabular}{lccc}
\toprule
Quantity & $R_{avg} = 0.1$ kpc & $R_{avg} = 0.5$ kpc & $R_{avg} = 1.0$ kpc \\
\midrule
\multicolumn{4}{c}{\textbf{Model T}} \\
\midrule
$B_{\mathrm{tot}}$ & $-2.49 \pm 0.06$ & $-2.49 \pm 0.06$ & $-2.49 \pm 0.06$ \\
$B_{0}$            & $-2.86 \pm 0.05$ & $-5.16 \pm 0.19$ & $-5.57 \pm 0.11$ \\
$\delta B$         & $-1.13 \pm 0.03$ & $-2.17 \pm 0.08$ & $-2.47 \pm 0.06$ \\
$v_{\mathrm{tot}}$ & $-3.83 \pm 0.06$ & $-3.83 \pm 0.06$ & $-3.83 \pm 0.06$ \\
$v_{0}$            & $-4.13 \pm 0.08$ & $-5.23 \pm 0.12$ & $-4.65 \pm 0.11$ \\
$\delta v$         & $-1.93 \pm 0.03$ & $-2.90 \pm 0.08$ & $-3.82 \pm 0.14$ \\
\midrule
\multicolumn{4}{c}{\textbf{Model R}} \\
\midrule
$B_{\mathrm{tot}}$ & $-2.68 \pm 0.05$ & $-2.68 \pm 0.05$ & $-2.68 \pm 0.05$ \\
$B_{0}$            & $-3.39 \pm 0.07$ & $-5.70 \pm 0.20$ & $-6.16 \pm 0.07$ \\
$\delta B$         & $-0.91 \pm 0.05$ & $-2.21 \pm 0.03$ & $-2.81 \pm 0.06$ \\
$v_{\mathrm{tot}}$ & $-3.38 \pm 0.07$ & $-3.38 \pm 0.07$ & $-3.38 \pm 0.07$ \\
$v_{0}$            & $-3.36 \pm 0.06$ & $-5.44 \pm 0.13$ & $-4.84 \pm 0.12$ \\
$\delta v$         & $-1.66 \pm 0.03$ & $-2.26 \pm 0.06$ & $-3.10 \pm 0.07$ \\
\bottomrule
\end{tabular}
\end{table}

\subsection{Time evolution of the energy densities}

\begin{figure*}
    \centering
    \includegraphics[trim={0.65cm 0cm 1.9cm 0cm},clip,width=.49\linewidth]{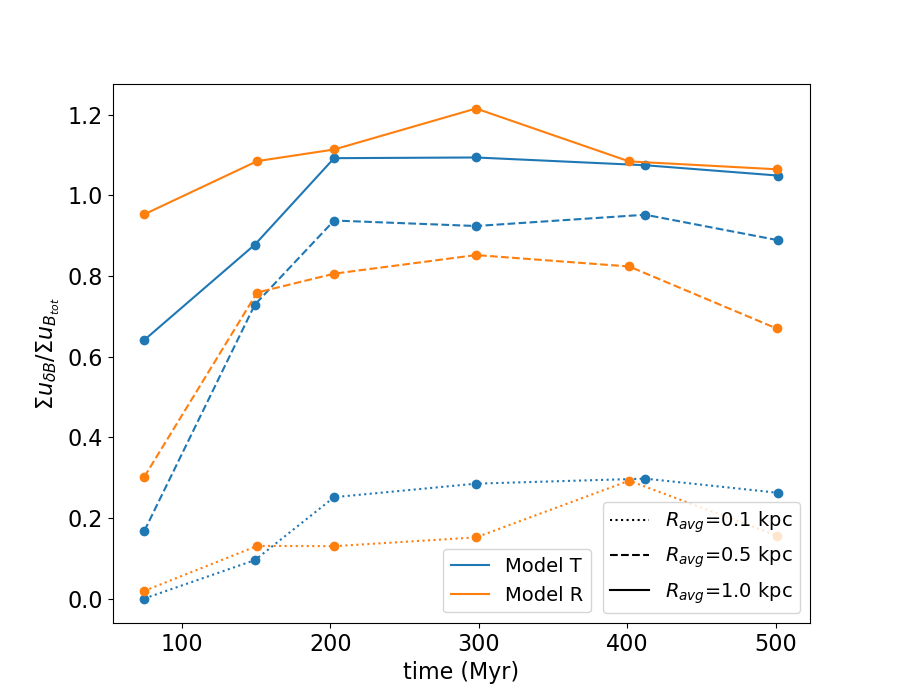} 
    \includegraphics[trim={0.5cm 0cm 1.9cm 0cm},clip,width=.5\linewidth]{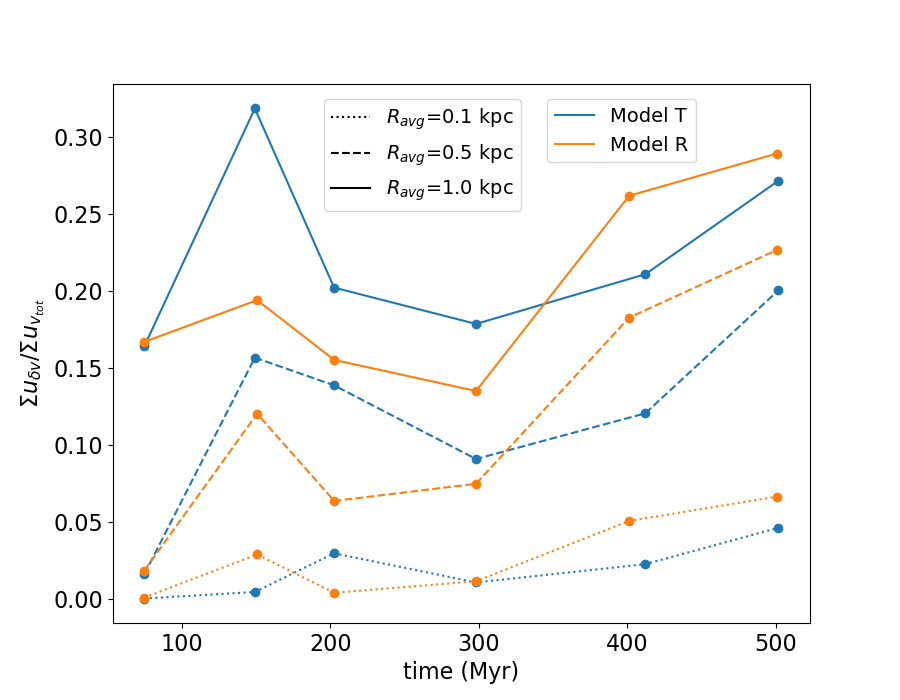}\\
    \caption{Time evolution of the ratio of turbulent to total energy for the magnetic and kinetic components in Models T and R, shown for different averaging radii.}
    \label{fig:ratios_t}
\end{figure*}

\begin{figure}
    \centering
    \includegraphics[trim={0.5cm 0cm 1.9cm 0cm},clip,width=.99\linewidth]{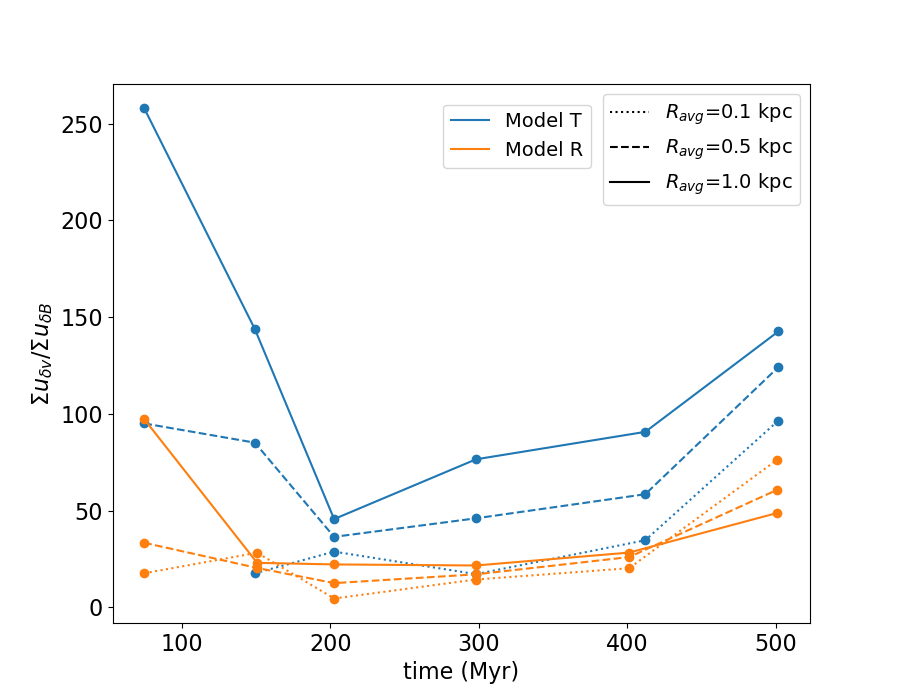}
    \caption{Time evolution  of the turbulent kinetic to magnetic energy, for models T and R, shown for different averaging radii.}
    \label{fig:dvdB_t}
\end{figure}

Fig.~\ref{fig:ratios_t} shows the ratio of turbulent to total magnetic energy (left) and turbulent to total kinetic energy (right) as a function of time. The most prominent feature in both cases is the strong dependence on the averaging radius. Larger radii yield systematically higher turbulent fractions, as the mean field becomes smoother, and a greater portion of the fluctuations is classified as turbulent. At a given time (e.g., 150 Myr), the inferred turbulent fraction of the kinetic energy density can vary from below 5\% to nearly 30\%, depending solely on the averaging scale. This suggests that the apparent level of turbulence is not an intrinsic property of the system, but rather a scale-dependent outcome of the chosen averaging procedure.

Within this framework, the two models follow similar overall trends. The turbulent magnetic energy initially grows and then stabilizes, a trend particularly pronounced for Model T. Model R consistently shows higher turbulent-to-total ratios than Model T during the early stages, reflecting its more disordered initial conditions. However, after 150 Myr, Model T surpasses Model R for the $R_{avg}$=0.1 kpc and $R_{avg}$=0.5 kpc cases, while for $R_{avg}$=1.0 kpc, Model R dominates at all times. Interestingly, for $R_{\text{avg}} = 1.0$ kpc, the turbulent magnetic energy density slightly exceeds the total magnetic energy density. This behavior can be understood from the energy decomposition,
$u_{Btot}=u_{B_0}+u_{\delta B}+2B_0\cdot \delta B/8\pi$ , where the cross term $B_0\cdot \delta B$ can be negative. At large averaging scales, the mean field becomes very smooth while the residual fluctuations may remain partially correlated but oppositely directed, producing a negative cross term. This reduces the total magnetic energy relative to the sum of the mean and turbulent components, which can lead to ratios greater than unity. \footnote{We have verified that in such cases the cross term is indeed negative, and the sum of $u_{B_0}$, $u_{\delta B}$, and $\frac{2B_0\cdot \delta B}{8\pi}$ exactly reproduces $u_{B_{\text{tot}}}$, confirming that the total magnetic energy derived from the simulations is consistent with the decomposition.}
For the kinetic energy fraction, a similar trend is observed with the averaging scale, where larger radii lead to higher turbulent fractions. Model T dominates Model R for $R_{avg}$=0.5 and 1.0 kpc during the early stages (up to 300 Myr), while at later times Model R overtakes Model T. The strong variation with scale demonstrates that even modest changes in the averaging radius can substantially modify the inferred turbulent energy content, emphasizing the need for caution when comparing results based on different spatial averaging scales.

Fig.~\ref{fig:dvdB_t} shows the ratio of turbulent kinetic and magnetic energy, which offers a complementary view, directly probing the balance between turbulent gas motions and magnetic turbulence. In both models, the turbulent kinetic energy exceeds the turbulent magnetic energy at all times, with Model T generally exhibiting larger ratios. This behavior is consistent with the slightly stronger magnetic field amplification in Model R discussed above, which leads to a relatively lower kinetic-to-magnetic energy ratio.
It is worth noting that, in general, larger averaging radii yield higher ratios. However, for Model R at 500 Myr, the opposite trend appears where the ratio decreases with increasing radius. This inversion arises because the magnetic turbulence grows more strongly with the averaging radius, leading to a relative reduction in the kinetic-to-magnetic turbulent energy ratio. This interpretation is consistent with the power spectra in Fig.~\ref{fig:ps_eres_500}, where the residual energy is significantly higher at $R_{\text{avg}}=0.1$ kpc for Model R.

\section{Discussion}\label{sec:discuss}
The results presented above reveal clear averaging scale-dependent behavior in both the magnetic and kinetic turbulence, which we now interpret in terms of turbulence characterization, the implications for observational diagnostics, and the imprint of initial conditions.

\subsection{Implications for magnetic field estimates and turbulence characterization}
The strong dependence of the turbulent energy fractions on the averaging scale has important implications for observational methods, such as the Davis-Chandrasekhar-Fermi (DCF) technique. 
The DCF method estimates the magnetic field strength from measurements of the gas density, the velocity dispersion, and the dispersion of polarization angles, under the assumption that turbulent kinetic and magnetic energies are in approximate equipartition.
However, two important issues affect the validity of this approach.
First, the assumption of equipartition is often violated. Only the idealized case of perfectly linear Alfv\'enic turbulence has no residual energy. A net negative residual energy always arises in realistic setups even for pure Alfv\'enic turbulence due to non-linear interactions \citep[e.g.,][]{muller2005,wang11}. A positive residual energy is usually the result of compressibility in super-Alfv\'enic turbulence \citep[e.g.,][]{lim20} and even in sub-Alfv\'enic \citep{skalidis2021}.
The power spectra of the residual energy in our models further demonstrate that this imbalance depends on spatial scale, highlighting that the equipartition assumption of DCF is not universally valid. This issue has been discussed in detail by \citet{ST2021, SSBPT2021, STP2023}.

The second issue concerns the sensitivity of the velocity dispersion and the dispersion of polarization angles to the chosen averaging scale. Since the turbulent fraction varies substantially with the averaging radius, the choice of scale can significantly affect DCF-based estimates of the magnetic field strength. Using a small averaging radius may underestimate the turbulent contribution, leading to an overestimation of the field strength, while a larger radius may have the opposite effect.

This sensitivity on the averaging scale extends to the characterization of strong and weak turbulence. In the weak turbulence regime, magnetic perturbations are small compared to the mean field ($\delta B << B_0$), and nonlinear interactions between turbulent perturbations proceed slowly \citep{nazarenko11,Schekochihin}. When wave timescale becomes comparable to non-linear timescale ($\delta B k_\bot /B_0 k_\| \sim 1$), the system transitions to strong turbulence in which nonlinear processes dominate energy transfer between scales \citep{Goldreich, Schekochihin}. In both models, the turbulent component dominates at small spatial scales (smaller than the averaging radius), indicating a regime of strong turbulence where magnetic fluctuations are comparable to or exceed the mean field. At larger scales, the ordered component becomes dominant, corresponding to weak turbulence.

Our results are consistent with the findings of \cite{ntormousi2024}, who showed that both strong and weak turbulence regimes can coexist within the same galactic system, even when initialized with an ordered magnetic configuration. In their analysis of the same galaxy simulations, they applied a local filtering technique based on the K-D tree method \citep{Maneewongvatana} to decompose the magnetic field into mean and fluctuating components. Despite using a similar averaging method to this work, they did not explore the effect of the chosen averaging scale on the turbulence characterization.

Our analysis demonstrates that the apparent strength of turbulence is highly scale-dependent so the classification of turbulence as strong or weak depends sensitively on the averaging radius used to define the mean and turbulent components. This highlights that turbulence characterization in galactic disks cannot be considered intrinsic, but must always account for the spatial scale over which the decomposition is performed.

\subsection{Remarks on the density field}

In this work we have examined the effect of averaging on how the flow and magnetic field turbulence are  defined. In a turbulent flow, the density field is expected to have a similar behavior when subject to averaging. However, a detailed analysis of density fluctuations is beyond the scope of the present study.
In galactic disks, the density field develops a strongly hierarchical structure as a result of the interplay between supersonic turbulence and self-gravity, with bound overdensities embedded within larger-scale structures across a wide range of spatial scales \citep[e.g.,][]{elmegreen2004,federrath2013}.

In this context, \citet{louvet2021} showed that the relative contribution of large- and small-scale density fluctuations depends sensitively on the method used to identify structures and to account for gravitational binding, highlighting the intrinsic scale dependence of density statistics.
This difference is further amplified by the fact that velocity and magnetic fields are vector quantities, whose scale dependence is influenced by directional correlations and anisotropies, whereas density is a scalar field whose structure is primarily shaped by compression and self-gravity.

\subsection{Initial conditions of magnetic field}
Examining the time evolution of magnetic energy densities, we see that after star formation is activated in the simulations (at 100 Myr), the magnetic fields in both models reach a quasi-steady configuration, and the differences between them become relatively small. Nevertheless, the power spectra at 500 Myr still show a subtle imprint of the initial magnetic field in Model T, visible as a broad enhancement in the turbulent component at scales comparable to the disk. Compared to the effects of the averaging scale, however, the influence of the initial conditions is much less pronounced, indicating that the spatial averaging and decomposition method play a far larger role in shaping the observed turbulent and mean-field properties.

\section{Conclusions}\label{sec:conclus}
In this work, we analyzed two Milky-Way-sized disk galaxy simulations with different initial magnetic morphologies, Model T with a toroidal field and Model R with a random field, using a spherical filtering method to decompose the magnetic and velocity fields into mean and turbulent components. Our results reveal that the characterization of magnetized turbulence in galactic disks depends on the averaging scale.

\begin{enumerate}
    \item The decomposition of magnetic and velocity fields into mean and turbulent components is strongly dependent on the spatial averaging scale. Larger smoothing scales yield higher apparent turbulence fractions as more large-scale structure is included in the fluctuating component.
    \item The turbulent kinetic energy dominates over the magnetic counterpart in both models, remaining one to two orders of magnitude higher in absolute terms. Nevertheless, the turbulent magnetic energy constitutes a dynamically significant fraction of the total magnetic energy. The relative contributions of kinetic and magnetic turbulence, however, depend strongly on the averaging scale where larger smoothing radii yield higher turbulent fractions in both fields.
    \item Although initial magnetic field configurations leave detectable imprints, such as a broad enhancement in the turbulent magnetic spectrum of Model T, these effects are secondary compared to the influence of the averaging scale.
    \item At small scales, the turbulent magnetic component dominates, corresponding to strong turbulence where fluctuations are comparable to or exceed the mean field. At larger scales, the ordered component prevails, representing a weakly turbulent regime. However, this distinction is inherently scale-dependent, emphasizing that the classification of turbulence as strong or weak depends on the spatial scale used in the field decomposition.
    \item The strong dependence of the turbulent energy fractions on the averaging scale directly affects magnetic field estimates derived from methods such as the DCF technique. The inferred field strength depends on the chosen averaging scale, emphasizing the need for consistent definitions between simulations and observations.    
\end{enumerate}

Overall, this study shows that the ordered and turbulent energy, both magnetic and kinetic, and the resulting classification of turbulence as strong or weak are scale-dependent outcomes of the chosen averaging or filtering method. Consequently, careful consideration of the averaging scale is essential not only for interpreting galactic magnetic fields, but also for analyzing the structure of interstellar turbulence in both simulations and observations.

\begin{acknowledgements}

AK and EN acknowledge funding from the Italian Ministry for Universities and Research (MUR) through the “Young Researchers” funding call (Project MSCA 000074).
    K.T. acknowledges the support by the TITAN ERA Chair project (contract no. 101086741) within the Horizon Europe Framework Program of the European Commission.
    This work was supported by computational time granted from the National Infrastructures for Research and Technology S.A. (GRNET S.A.) in the National HPC facility - ARIS - under project ID pr013007$\_$thin. 
    Part of the work has been performed under the Project HPC-EUROPA3 (INFRAIA-2016-1-730897), with the support of the EC Research Innovation Action under the H2020 Programme.
    We also gratefully acknowledge the computational resources of the Center for High Performance Computing (CHPC) at the University of Crete.
    We acknowledge usage of the Python programming language \citep{python2,python3}, Matplotlib \citep{matplotlib}, NumPy \citep{numpy}, and SciPy \citep{scipy}.
\end{acknowledgements}


\bibliographystyle{aa}
\bibliography{citations}

\begin{appendix}

\section{Test of the spherical filtering method}\label{appendix:a}
To test the validity of our method, we first constructed a purely toroidal magnetic field and applied our spherical averaging procedure. In this case, the mean field is identical to the total field at all averaging radii, and the turbulent component vanishes, as expected. We then introduced a 20$\%$ random perturbation to the toroidal field. The resulting histograms show that the ratio $|\delta B|/|B_0|$ increases, with a mean value of $\sim 0.2$, consistent with the imposed level of randomness.

\begin{figure}[!hb]
    \centering
    \hspace{0.2cm} \large Purely toroidal  \\
    \includegraphics[trim={0cm 0cm 0cm 0cm},clip,width=.99\linewidth]{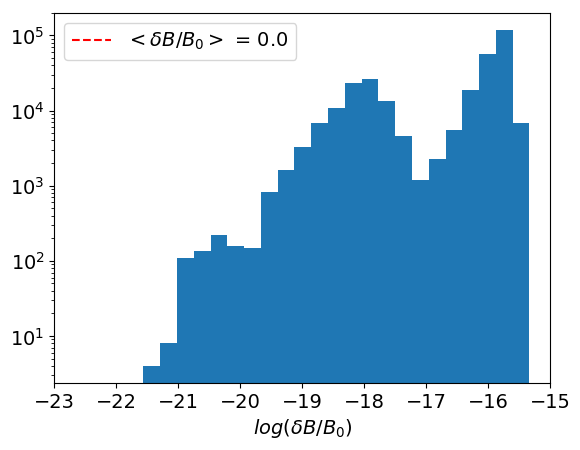} 
    \hspace{5.4cm} \large 20$\%$ randomness 
    \includegraphics[trim={0.cm 0cm 0cm 0cm},clip,width=.99\linewidth]{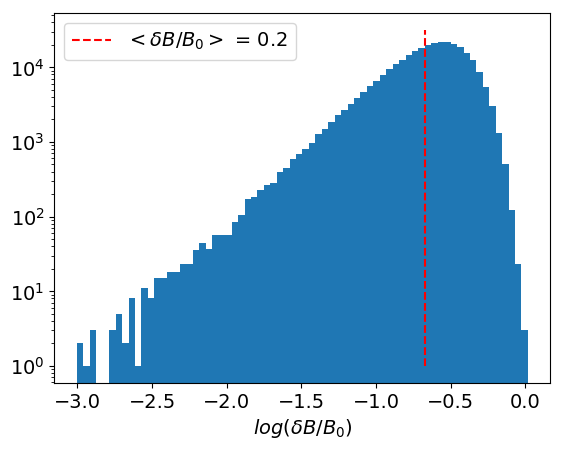}
    \caption{Histograms of $\log (\delta B / B_{0})$ for a purely toroidal field (left) and for a field with 20$\%$ perturbations added (right). The dashed red line indicates the mean value of $\delta B / B_{0}$. In the toroidal case, values remain close to machine precision, confirming $\delta B \approx 0$, whereas with 20$\%$ perturbations the mean shifts to $\simeq 0.2$, as expected from the imposed randomness.}
    \label{fig:hist_test}
\end{figure}

\begin{figure*}[!hb]
    \centering
    \hspace{1.5cm} \large Purely toroidal  \hspace{5.3cm} \large 20$\%$ randomness \\
    \rotatebox{90}{\hspace{2.8cm} \Large$B_{tot}$}
    \includegraphics[trim={0cm 0cm 0cm 0cm},clip,width=.44\linewidth]{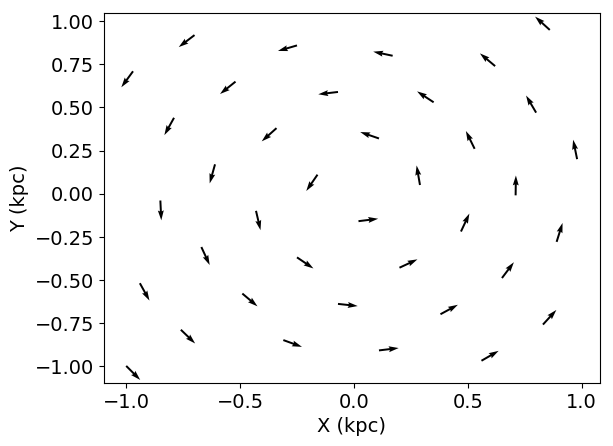} 
    \includegraphics[trim={0.cm 0cm 0cm 0cm},clip,width=.44\linewidth]{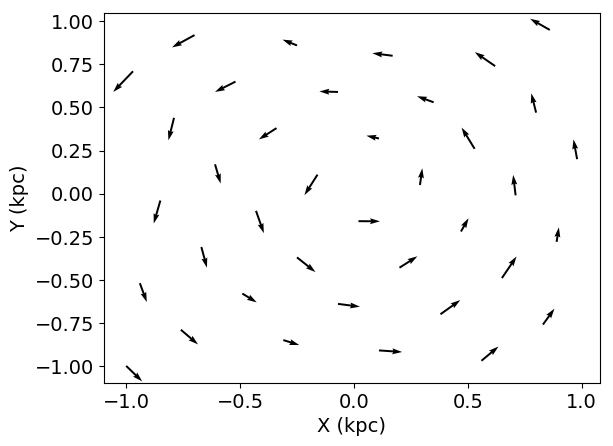}\\
    \rotatebox{90}{\hspace{2.8cm} \Large$B_{0}$}
    \includegraphics[trim={0cm 0.cm 0cm 0cm},clip,width=.44\linewidth]{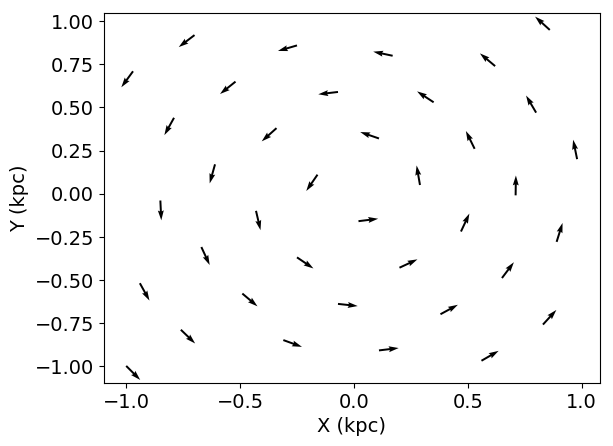} 
    \includegraphics[trim={0cm 0cm 0cm 0cm},clip,width=.44\linewidth]{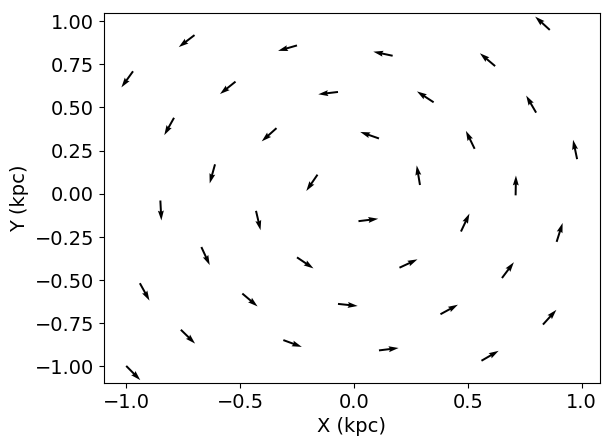}\\
    \rotatebox{90}{\hspace{2.8cm} \Large$\delta B$}
    \includegraphics[trim={0cm 0cm 0cm 0cm},clip,width=.44\linewidth]{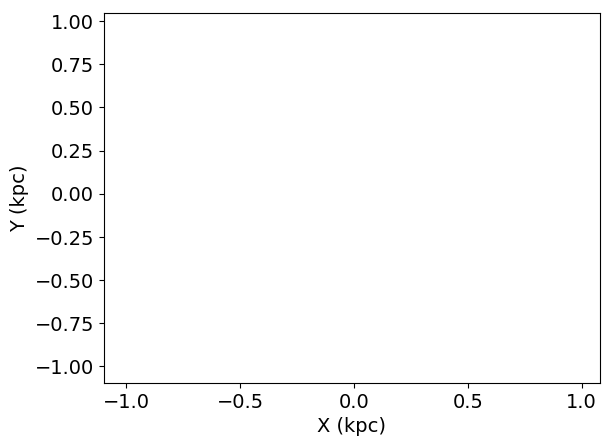} 
    \includegraphics[trim={0cm 0cm 0cm 0cm},clip,width=.44\linewidth]{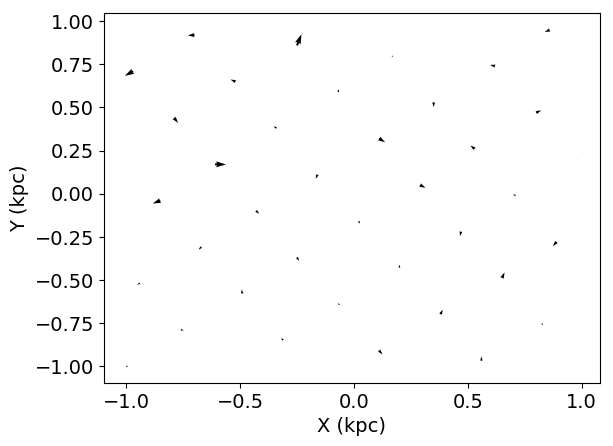}
    \caption{Test of the spherical averaging (filtering) method. Maps of the total magnetic field ($B_{\mathrm{tot}}$), the mean field ($B_{0}$), and the turbulent component ($\delta B$) are shown for a purely toroidal configuration (left panels) and for the same field with 20$\%$ added randomness (right panels). In the toroidal case, $B_{0}$ coincides with $B_{\mathrm{tot}}$ and $\delta B \approx 0$, while the randomized case introduces a finite turbulent contribution. }
    \label{fig:B_maps_test}
\end{figure*}

\newpage
\section{Power spectra of magnetic and velocity fields}\label{appendix:b}
\begin{figure*}
    \centering
    \hspace{1.3cm}\large Model T  \hspace{7.3cm} \large Model R \\
    \rotatebox{90}{\hspace{2.6cm} \large $R_{avg}$=0.1 kpc}
    \includegraphics[trim={0cm 0cm 0.3cm 0cm},clip,width=.48\linewidth]{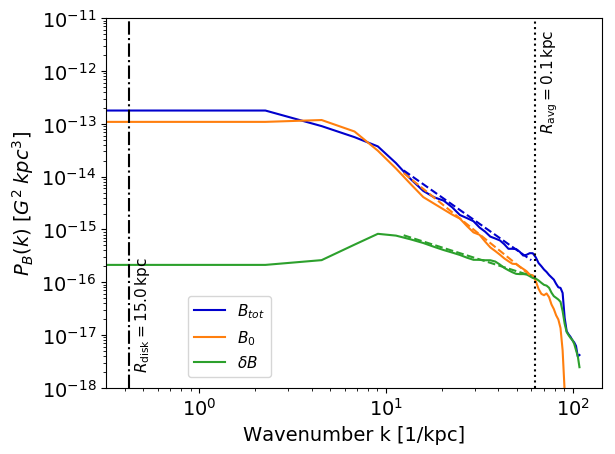} 
    \includegraphics[trim={0.78cm 0cm 0cm 0cm},clip,width=.47\linewidth]{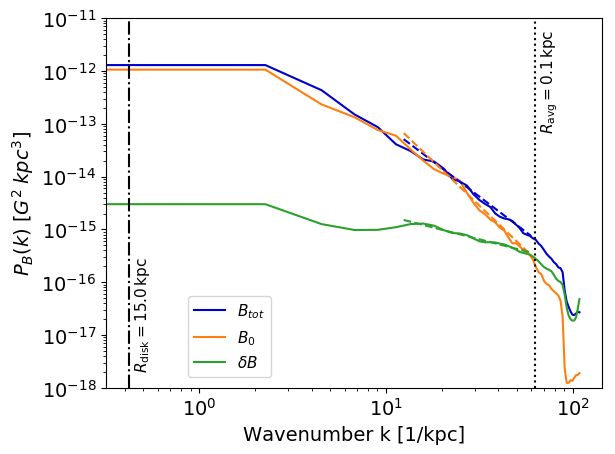}\\
    \rotatebox{90}{\hspace{2.6cm} \large $R_{avg}$=0.5 kpc}
    \includegraphics[trim={0cm 0.cm 0.3cm 0cm},clip,width=.48\linewidth]{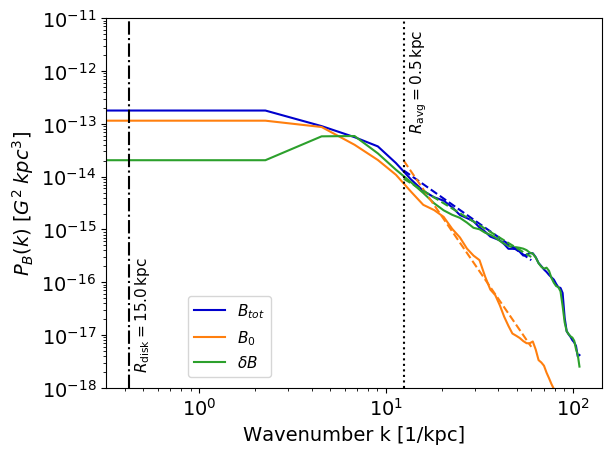} 
    \includegraphics[trim={0.78cm 0cm 0cm 0cm},clip,width=.47\linewidth]{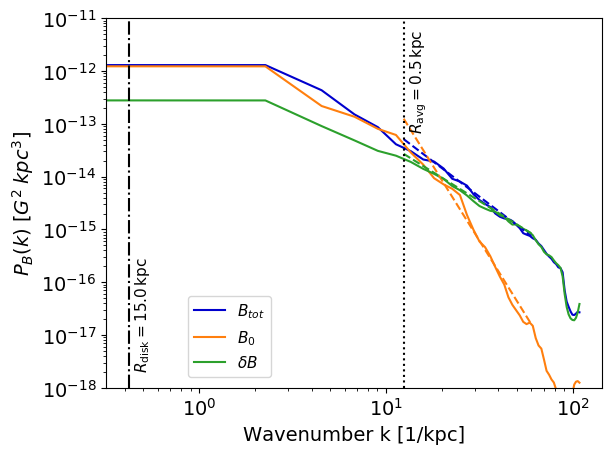}\\
    \rotatebox{90}{\hspace{2.6cm} \large $R_{avg}$=1.0 kpc}
    \includegraphics[trim={0cm 0cm 0.3cm 0cm},clip,width=.48\linewidth]{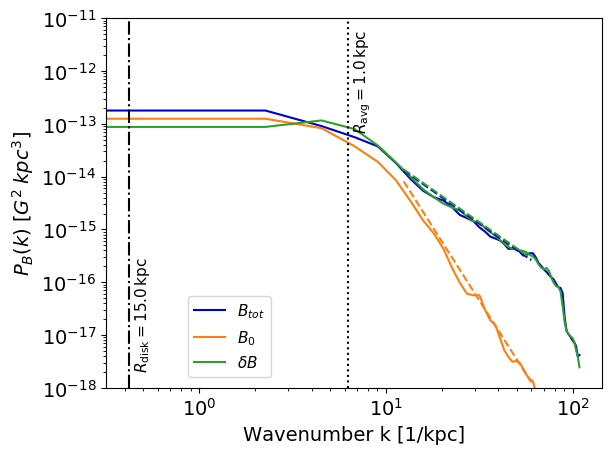} 
    \includegraphics[trim={0.78cm 0cm 0cm 0cm},clip,width=.47\linewidth]{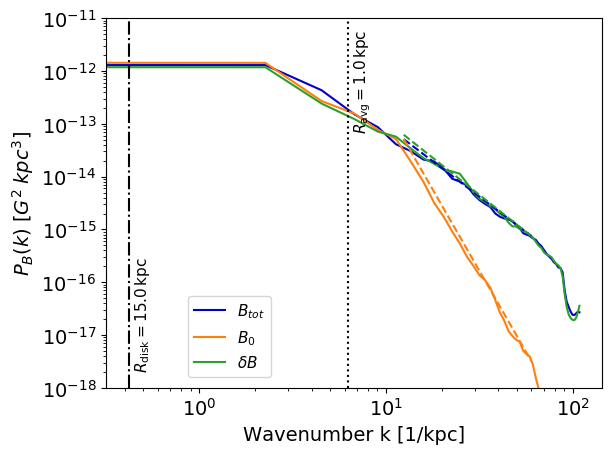}
    \caption{Power spectra of the total field ($P_{B_{tot}}$), the mean field ($P_{B_0}$) and the turbulent component ($P_{\delta B}$) at 500 Myrs. Dashed lines indicate the fits. Model T and model R are shown on the left and right side, respectively. }
    \label{fig:ps_B_500}
\end{figure*}

Fig~ \ref{fig:ps_B_500} shows the power spectra of the total magnetic field $B_{tot}$, the mean field $B_0$, and the fluctuating component $\delta B$, calculated for different averaging radii. The same trends discussed in Sec.~\ref{sec:results} are apparent. The fitted power laws are also displayed. Importantly, for spatial scales smaller than $R_{\rm avg}$, the power in the turbulent magnetic component exceeds that of the mean field, reflecting the dominance of magnetic fluctuations below the filtering scale.

\begin{figure*}
    \centering
    \hspace{1.3cm}\large Model T  \hspace{7.3cm} \large Model R \\
    \rotatebox{90}{\hspace{2.7cm} \large $R_{avg}$=0.1 kpc}
    \includegraphics[trim={0cm 0cm 0.3cm 0cm},clip,width=.48\linewidth]{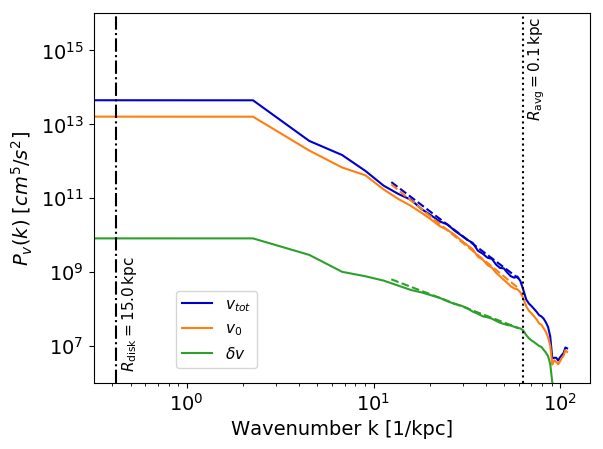} 
    \includegraphics[trim={0.78cm 0cm 0cm 0cm},clip,width=.47\linewidth]{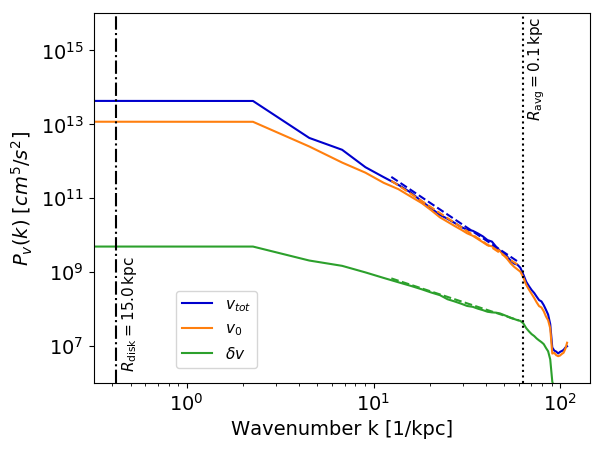}\\
    \rotatebox{90}{\hspace{2.7cm} \large $R_{avg}$=0.5 kpc}
    \includegraphics[trim={0cm 0.cm 0.3cm 0cm},clip,width=.48\linewidth]{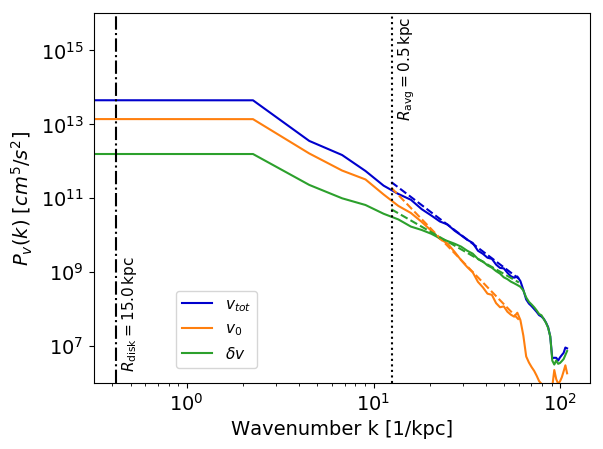} 
    \includegraphics[trim={0.78cm 0cm 0cm 0cm},clip,width=.47\linewidth]{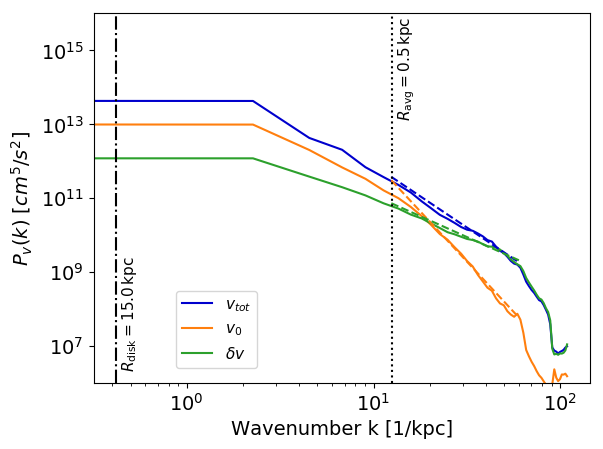}\\
    \rotatebox{90}{\hspace{2.7cm} \large $R_{avg}$=1.0 kpc}
    \includegraphics[trim={0cm 0cm 0.3cm 0cm},clip,width=.48\linewidth]{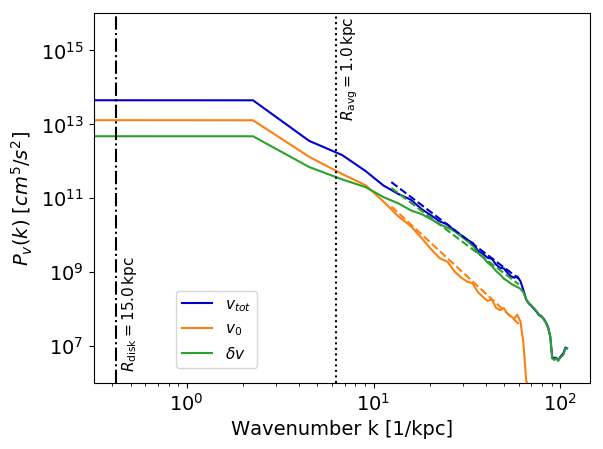} 
    \includegraphics[trim={0.78cm 0cm 0cm 0cm},clip,width=.47\linewidth]{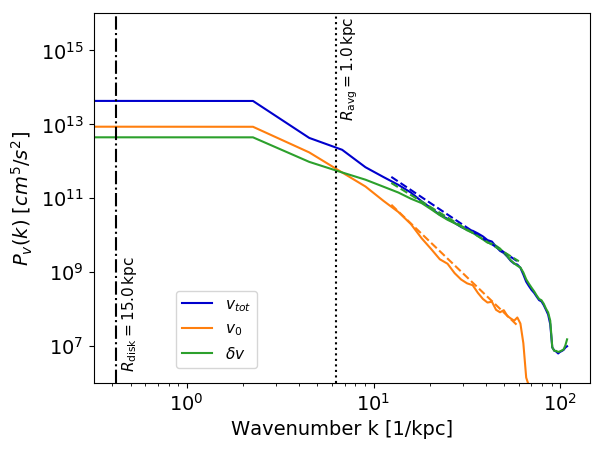}
    \caption{Power spectra of the total field ($P_{v_{tot}}$), the mean field ($P_{v_0}$) and the turbulent component ($P_{\delta v}$) at 500 Myrs. Dashed lines indicate the fits. Model T and model R are shown on the left and right side, respectively. }
    \label{fig:ps_v_500}
\end{figure*}

Fig~ \ref{fig:ps_v_500} shows the corresponding power spectra for velocity fields. Similar trends are observed as in the magnetic field analysis.
Likewise, for scales smaller than $R_{avg}$, the turbulent velocity component dominates over the mean component.
In both cases, the separation between the mean and turbulent spectra occurs around the wavenumber associated with the chosen averaging radius, as expected from the filtering procedure. This highlights that the inferred balance between ordered and turbulent components depends explicitly on the adopted averaging scale.

\end{appendix}

\end{document}